\documentclass[a4paper]{jpconf}
\usepackage[pdftex]{graphicx}
\usepackage{amssymb}
\usepackage{aas_macros}
\usepackage{url}

\def\farcs{\hbox{$.\!\!^{\prime\prime}$}}

\begin{document}
\title{The SED of Low-Luminosity AGNs at high-spatial resolution}


\author{J~A Fern\'andez-Ontiveros$^{1,2,3}$, M~A Prieto$^{1,2}$, J~A Acosta-Pulido$^{1,2}$ and M Montes$^{1,2}$}

\address{$^1$ Max-Planck-Institut f\"ur Radioastronomie (MPIfR), Auf dem H\"ugel 69, Bonn, D--53121, Germany}
\address{$^2$ Instituto de Astrof\'isica de Canarias (IAC), V\'ia L\'actea s/n, La Laguna, E--38200, Spain}
\address{$^3$ Departamento de Astrof\'isica, Facultad de F\'isica, Universidad de La Laguna, Astrof\'isico Fco. S\'anchez s/n, La Laguna, E--38207, Spain}

\ead{jafo@mpifr.de}

\begin{abstract}

The inner structure of AGNs is expected to change below a certain luminosity limit. The big blue bump, footprint of the accretion disk, is absent for the majority of low-luminosity AGNs (LLAGNs). Moreover, recent simulations suggest that the torus, a keystone in the Unified Model, vanishes for nuclei with $L_{bol} \lesssim 10^{42}\, \rm{erg\, s^{-1}}$. However, the study of LLAGN is a complex task due to the contribution of the host galaxy, which light swamps these faint nuclei. This is specially critical in the IR range, at the maximum of the torus emission, due to the contribution of the old stellar population and/or dust in the nuclear region. Adaptive optics imaging in the NIR (VLT/NaCo) together with diffraction limited imaging in the mid-IR (VLT/VISIR) permit us to isolate the nuclear emission for some of the nearest LLAGNs in the Southern Hemisphere. These data were extended to the optical/UV range (\emph{HST}), radio (VLA, VLBI) and X-rays (\emph{Chandra}, \emph{XMM}-Newton, \emph{Integral}), in order to build a genuine spectral energy distribution (SED) for each AGN with a consistent spatial resolution ($< 0\farcs5$) across the whole spectral range. From the individual SEDs, we construct an average SED for LLAGNs sampled in all the wavebands mentioned before. Compared with previous multiwavelength studies of LLAGNs, this work covers the mid-IR and NIR ranges with high-spatial resolution data. The LLAGNs in the sample present a large diversity in terms of SED shapes. Some of them are very well described by a self-absorbed synchrotron (e.g. NGC~1052), while some other present a thermal-like bump at $\sim 1\, \rm{\mu m}$ (NGC~4594). All of them are significantly different when compared with bright Seyferts and quasars, suggesting that the inner structure of AGNs (i.e. the torus and the accretion disk) suffers intrinsic changes at low luminosities.
\end{abstract}

\section{Introduction}

The majority of AGNs spend their lives in a low state, characterized by a low accretion rate and a modest luminosity. These are known as low-luminosity AGNs (LLAGNs), the faintest but also the most numerous members of this family, including $\sim 1/3$ of all galaxies in the Local Universe \cite{2008ARA&A..46..475H}. However, LLAGNs are not just scale-down versions of the bright counterparts, Seyfert galaxies and quasars. Their main characteristics are: \textit{i)} absence of the \textbf{big blue bump} \cite{1996PASP..108..637H}, footprint of the accretion disk in the Spectral Energy Distribution (SED), \textit{ii)} LLAGNs are \textbf{``radio-loud''}, usually showing compact cores and parsec-scale radio jets \cite{2005A&A...435..521N}, and \textit{iii)} they are frequently associated with \textbf{low ionization} nuclear emission-line regions (LINERs, \cite{1980A&A....87..152H}). The differences between high-luminosity objects and LLAGNs might be associated with the structural changes predicted at low luminosities. Both the broad-line region and the ``torus'', keystones of the Unified Model \cite{1993ARA&A..31..473A}, are expected to vanish at $L_{bol} \lesssim 10^{42}\, \rm{erg/s}$ \cite{2003ApJ...590...86L,2007MNRAS.380.1172H}. Instead of the classical scheme, the nucleus of LLAGNs is described in terms of a \textbf{radiatively inefficient} structure in which most of the gravitational energy is not released in the form of electromagnetic radiation \cite{2005Ap&SS.300..177N,2008ARA&A..46..475H}. Its main components are: \textit{i)} a radiatively inefficient accretion flow (RIAF), \textit{ii)} a truncated accretion disk, and \textit{iii)} a jet or an outflow. These components contribute to the total energy output and compete at certain wavelength ranges, but the relative importance of each one of them is still not clear (e.g. \cite{2011ApJ...726...87Y}). Specifically, RIAF-dominated models predict that most of the energy is released at radio and X-ray wavelengths, while high-spatial resolution data suggest an important contribution in the IR (see Fig.~2). On the other hand, some LLAGNs still show broad lines in polarized light \cite{1999ApJ...525..673B} together with a high absorption column ($\gtrsim 10^{23}\, \rm{cm^{-2}}$, \cite{2009ApJ...704.1570G}), which claim for the presence of a torus in their nuclei. In brief, LLAGNs permit to explore the boundaries of unified model, the luminosity radiated is not high enough to support the torus and the structure of the accretion disk also seems to be altered at low accretion ratios.

\section{Dataset}\label{obs}

\begin{figure}
  \centering
  \includegraphics[width = 0.32\textwidth]{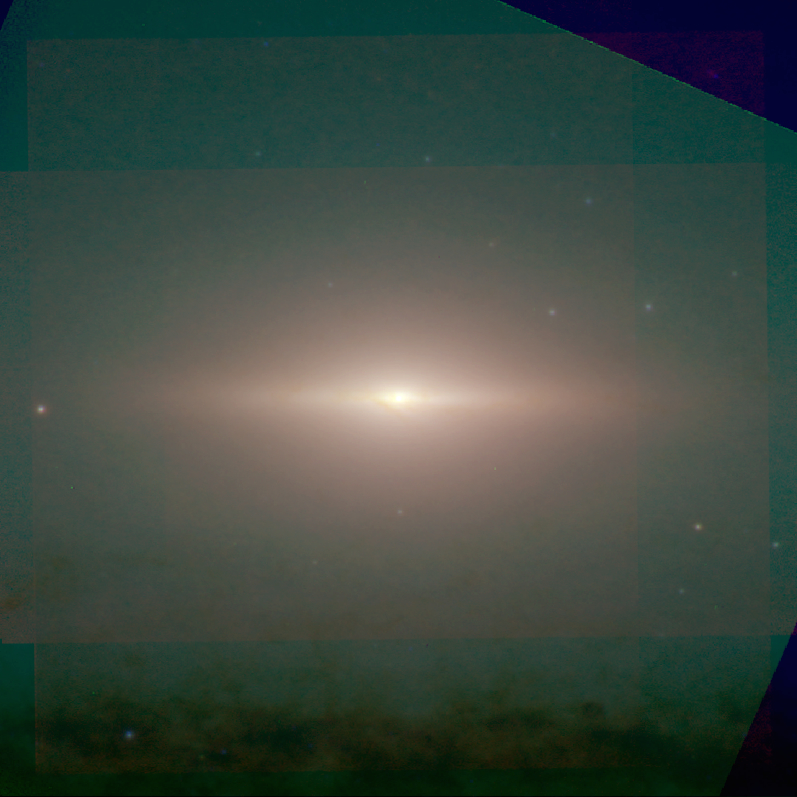}
  \includegraphics[width = 0.32\textwidth]{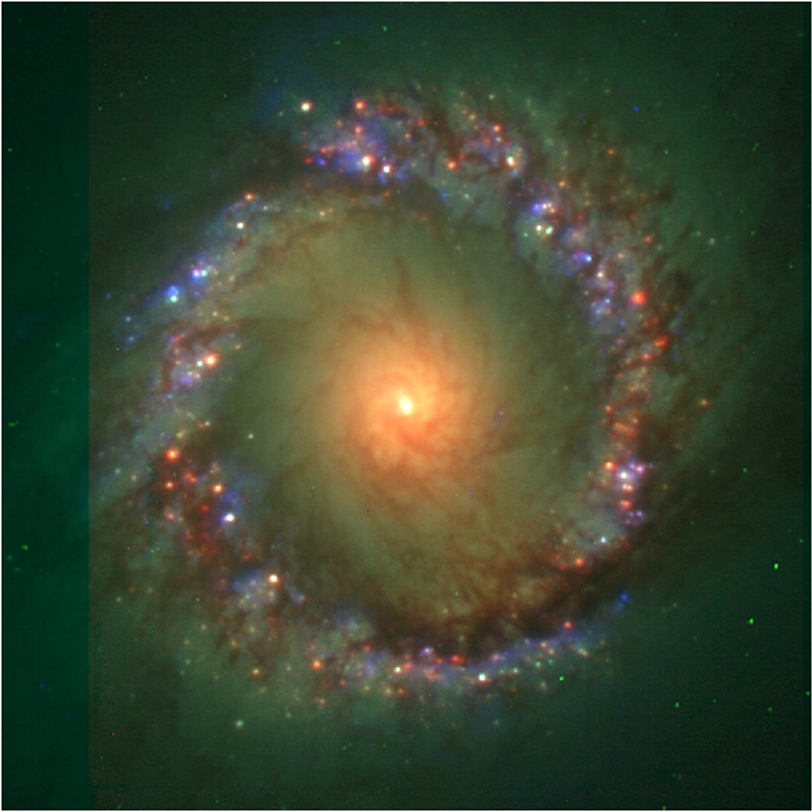}
  \includegraphics[width = 0.32\textwidth]{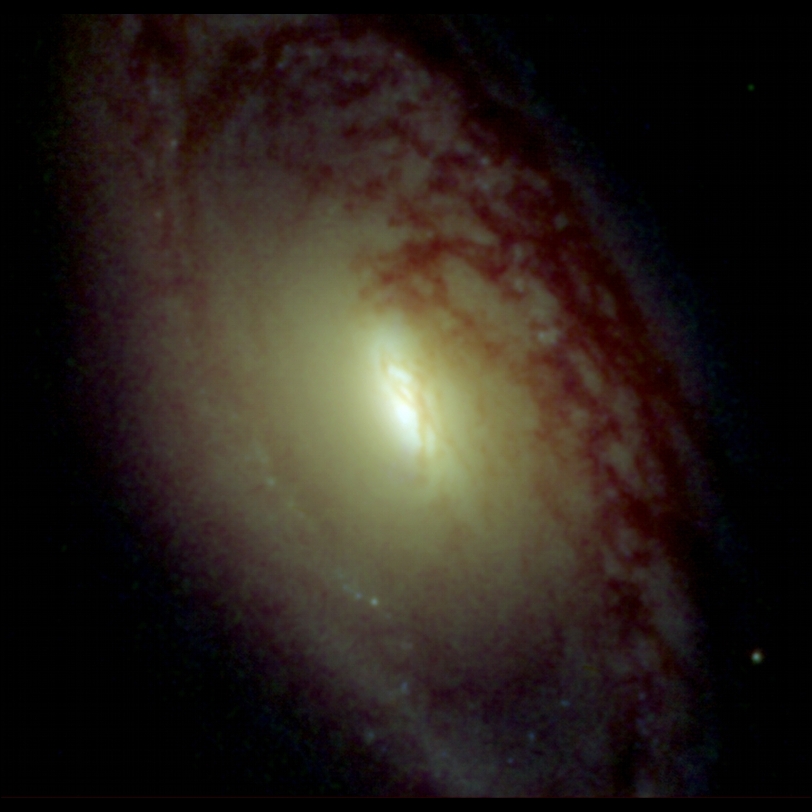}
  \includegraphics[width = 0.32\textwidth]{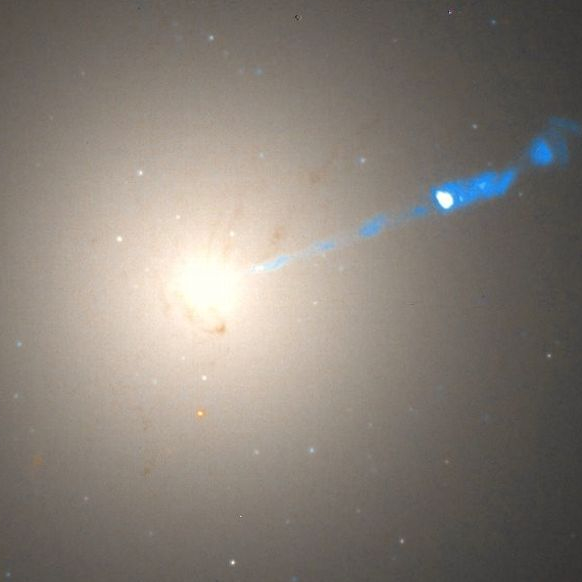}
  \includegraphics[width = 0.32\textwidth]{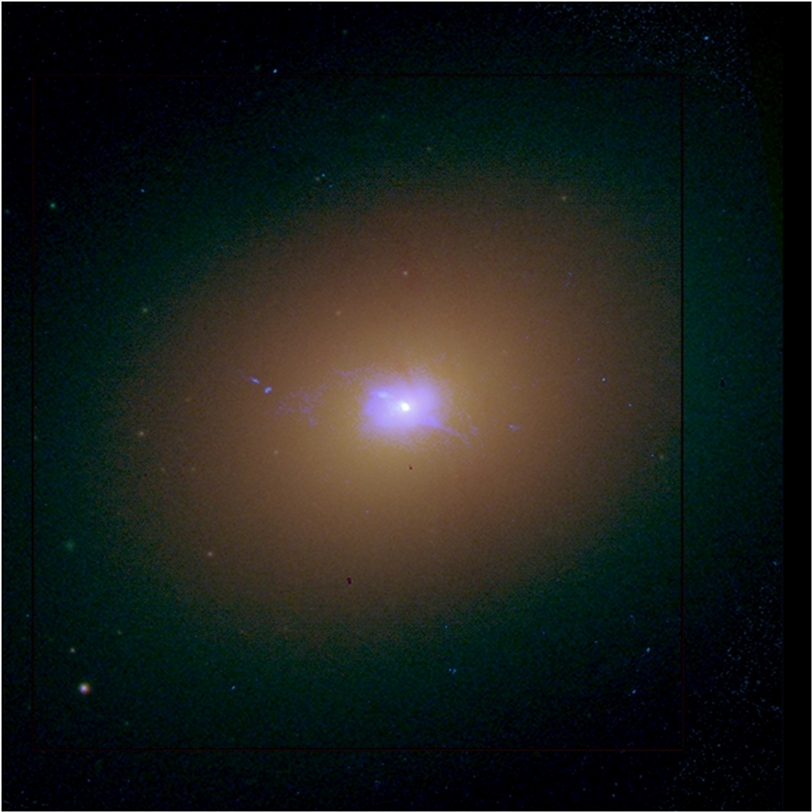}
  \includegraphics[width = 0.32\textwidth]{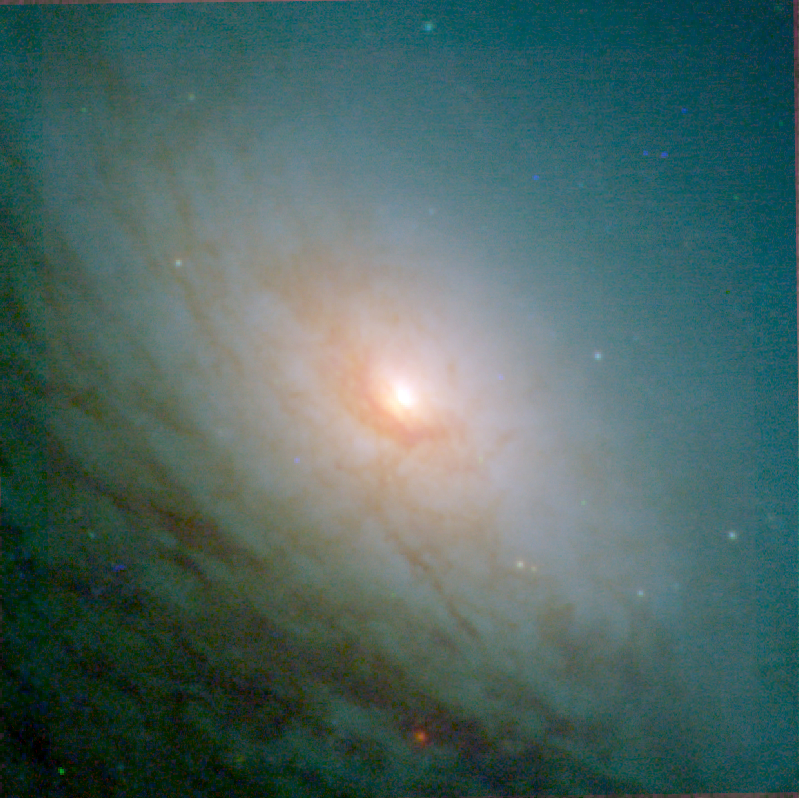}
  \caption{\label{im_sample} Optical-IR colour composition for the central parsecs of the galaxies in the sample. From left to right, top to bottom, the frames corresponds to NGC~4594 (FOV\,$= 15'' \times 15''$), NGC~1097 ($32'' \times 32''$), NGC~1386 ($26'' \times 26''$), M87 ($29'' \times 29''$), NGC~1052 ($30'' \times 30''$) and NGC~3169 ($10'' \times 10''$).}
\end{figure}

The dataset on which this work is based consist on a multiwavelength, high-spatial resolution dataset covering the radio, infrared, optical, UV and X-ray ranges. This project is a follow-up of \emph{The central parsecs of the nearest galaxies}\footnote{\url{http://www.iac.es/project/parsec}} \cite{2010MNRAS.402..879R,2010MNRAS.402..724P}, a high-spatial resolution study of the brightest and nearest Seyfert galaxies carried out at sub-arcsec scales with the Very Large Telescope (VLT). The objects included in the present work (Table~\ref{sample}, Fig.~\ref{im_sample}) correspond to AGNs one to two order of magnitude fainter ($L_{bol} \lesssim 10^{42}\, \rm{erg\, s^{-1}}$).

The observations include the near (NIR, VLT/NaCo) and mid-infrared (mid-IR, VLT/VISIR) ranges for the inner few kpc of a sample of 6 galaxies (Table~\ref{sample}). These observations were acquired using VLT/NaCo adaptive optics in the NIR and diffraction limited imaging with VLT/VISIR in the mid-IR. The dataset was completed with the highest-spatial resolution data available in the optical/UV range from the \emph{HST} scientific archive. Nuclear fluxes were measured using aperture photometry of the unresolved component in the centre ($\lesssim 0\farcs1$), subtracting the local background around ($0\farcs2$--$0\farcs3$). In the mid-IR, \emph{PSF} photometry was performed (see \cite{2010MNRAS.402..879R}). An extensive search in the literature has been performed in order to complete the radio and X-ray ranges. Flux measurements in these ranges correspond mainly to Very Large Array (VLA) and Very Long Baseline Interferometry (VLBI) in radio and \emph{Chandra}, \emph{XMM}-Newton and \emph{Integral} at X-rays. The characteristics of this dataset permit us to build, for each LLAGN in the sample, a consistent SED of the same physical region, very well sampled over a wide range in wavelength. This permit us to disentangle the nuclear emission from the host galaxy light, which otherwise dominates the energy output.
\begin{table}[h]
  \small
  \centering
  \caption[Sample]{Galaxy sample (objects sorted by increasing distance). \textsc{fwhm} correspond to the size of the most compact object found in the FOV. $\rm{L_{bol}}$ is estimated from the high-spatial resolution SEDs.
}\label{sample}
  \lineup
  \begin{tabular}{lcccccll}
    \br
    
\bf Name &
\bf D &
\bf Ref. &
\bf FWHM &
\bf 1$''$&
$\mathbf{L_{bol}}$ &
\bf Type &
\bf Class \cr

& [Mpc] & & [pc] & [pc] & [$\rm{erg\, s^{-1}}$] & & \cr
\mr
NGC 4594 & 9.08 & \cite{2003ApJ...583..712J} & 3.1  & \044.0 & $1.9 \times 10^{41}$ & SA(s)a & LINER 2 \cr
NGC 1097 & 14.2   & \cite{2009AJ....138..323T} & 8.9  & \068.8 & $1.9 \times 10^{42}$ & SB(r$'$l)b & LINER 1 \cr
NGC 1386 & 15.3   & \cite{2003ApJ...583..712J} & 6.7  & \074.2 & $6.6 \times 10^{42}$ & SB(s)a & Sy 2 \cr
M87      & 16.7   & \cite{2009ApJ...694..556B} & 9.2  & \081.0 & $1.2 \times 10^{42}$ & cD & LINER 1 \cr
NGC 1052 & 18.0   & \cite{2003ApJ...583..712J} & 10.5 & \087.3 & $9.3 \times 10^{42}$ & E4 & LINER 1.9 \cr
NGC 3169 & 24.7   & \cite{2009ApJ...704..629M} & 10.9 &  119.7 & $8.1 \times 10^{41}$ & SA(s)a & LINER 2 \cr

    \br
  \end{tabular}
\end{table}

\section{Multiwavelength SEDs}

Multiwavelength SEDs were built at a consistent subarsec resolution based on the aperture photometry data in the IR, optical and UV ranges together with the radio and X-ray measurements from the literature (Section~\ref{obs}). First, we will focus on the cases of NGC~1052 and the Sombrero galaxy (NGC~4594). Both are very well sampled in wavelength. The former is a very good example of a SED described by a power-law in the IR-to-UV range, while the latter present a different behaviour showing a thermal NIR bump at $\sim 1\, \rm{\mu m}$. The rest of the galaxies in the sample present a SED shape with some degree of similarity to one of these two cases.

\subsection{NGC 1052}
\begin{figure}
  \centering
  \includegraphics[width = 0.7\textwidth]{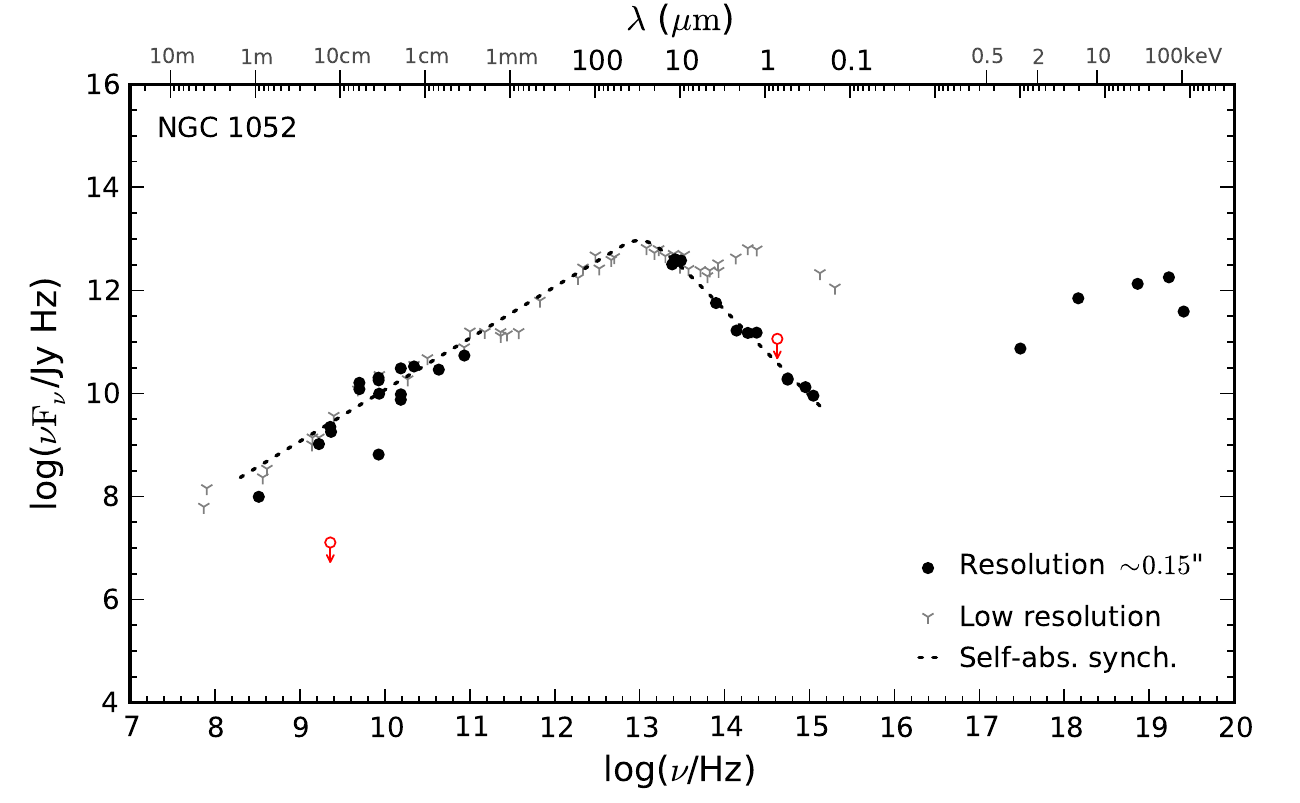}
  \caption{\label{n1052_sed} High-spatial resolution SED for NGC~1052 (black dots). Low-spatial resolution data are also shown for comparison (grey spikes). A self-absorbed synchrotron model, representative of a jet emission, has been fitted to the radio, IR, optical and UV range (dashed-line).}
\end{figure}

NGC~1052 is a prototypical example of a LINER type 2 nucleus \cite{1980A&A....87..152H}. Fig.~\ref{n1052_sed} shows the high-spatial resolution SED (black dots), together with large aperture measurements (grey spikes). The nucleus is the brightest source in the radio and mid-IR ranges, and seems to dominate the millimetre and sub-millimetre ranges as well. However, the bulk of the stellar light is nearly two order of magnitude above the nuclear emission for large apertures at optical and IR wavelengths. Only high-spatial resolution measurements permit to isolate the flux of the LLAGN and discern the nature of the emission in this range.

Based on the high X-ray absorption column ($N_H \sim 10^{23}\, \rm{cm^{-2}}$, \cite{2011ApJ...727...19F}) and the presence of a broad H$\alpha$ line in polarised light \cite{1999ApJ...515L..61B}, one would expect a strongly obscured nucleus for this galaxy in the optical/UV range. However, instead of a continuum affected by a high reddening, the high-spatial resolution SED follows a power law in the mid-IR-to-UV range (Fig.~\ref{n1052_sed}). Moreover, a variation in the position of the nucleus with wavelength, common in highly reddened AGNs, is not detected in this case. This power-law behaviour is similar to the case of NGC~4258 \cite{2000ApJ...531..756C}. A self-absorbed synchrotron \cite{2000A&A...361..850T}, i.e. a very simple model of a jet, describes very well the high-spatial resolution SED of NGC~1052 in the radio-to-UV range (dashed-line in Fig.~\ref{n1052_sed}). Radio core fluxes, corresponding to milliarcsec apertures are not included in the fit. The optically-thick part of the self-absorbed synchrotron, from radio to mid-IR, presents a flat slope ($\alpha \approx 0$, $S_\nu \propto \nu^{-\alpha}$), followed by a turnover and the optically-thin emission in the mid-IR-to-UV range ($\alpha \approx 2.7$). Although a detailed modelling is desired, this simple exercise show that the energy output in the nucleus of NGC~1052 is driven the jet. Other LLAGNs also show a similar behaviour (e.g. M81, \cite{2008ApJ...681..905M}). In this context, the X-ray flux would be associated to the comptonization of photons emitted by syncrotron radiation.

The IR range is described by the Unified Model \cite{1993ARA&A..31..473A} in terms of thermal emission from dust in a torus of a few parsecs in size. However, at a spatial resolution of $\sim 10.5\, \rm{pc}$, the dominant mechanism seems to be non-thermal. This is supported by the polarisation degree ($\sim 3$--$5\%$) found in the continuum emission by \cite{1982ApJ...252L..53R} in the IR and by \cite{1973ApJ...179L..93H} in the UV range, based on $7''$ and $10''$ diameter apertures, respectively. If a torus is present in the centre of this LLAGN, as suggested by the broad H$\alpha$ line in the polarised spectrum and the high $N_H$ column density, its contribution to the parsec-scale emission is very low.

\subsection{NGC 4594}
\begin{figure}
  \centering
  \includegraphics[width = 0.7\textwidth]{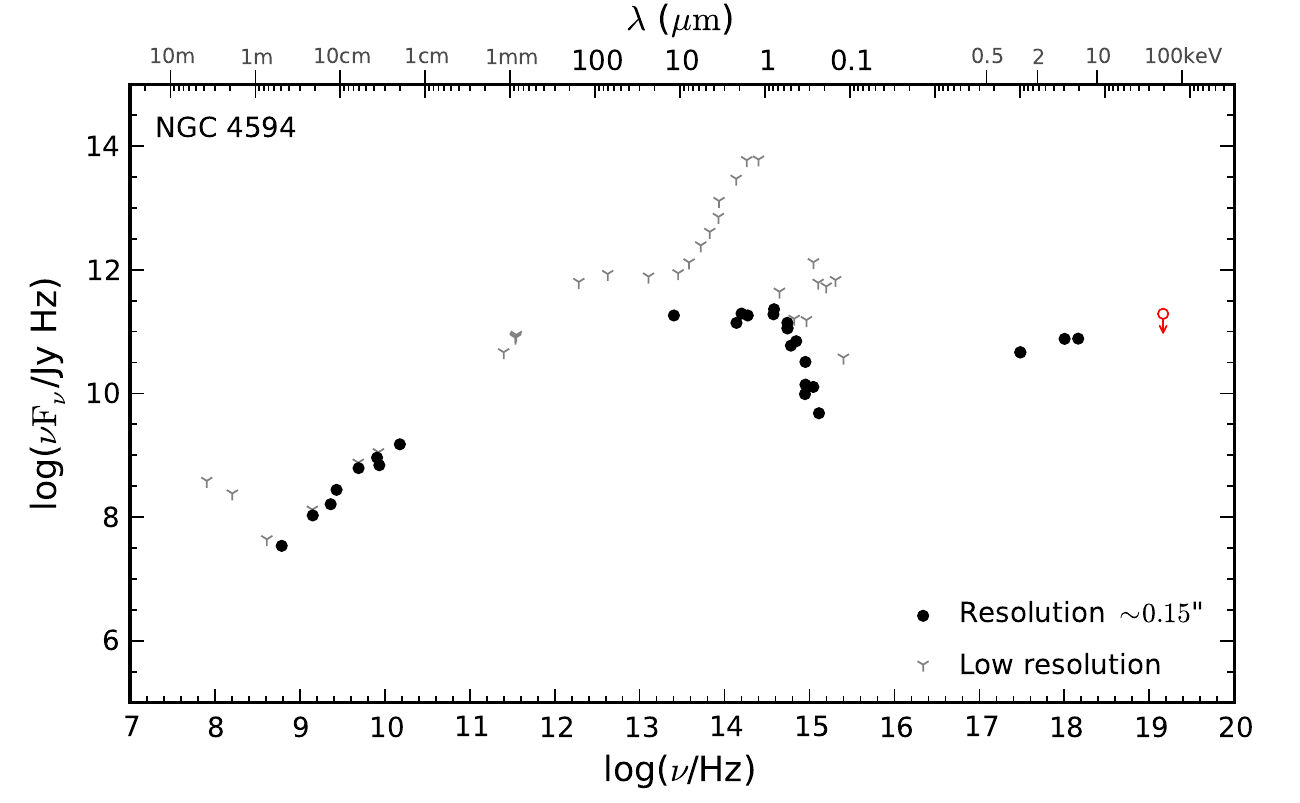}
  \caption{\label{n4594_sed} High-spatial resolution SED for NGC~4594 (black dots). Low-spatial resolution data are also shown for comparison (grey spikes).}
\end{figure}

Also known as the Sombrero galaxy or M104, NGC~4594 hosts the faintest nuclei in the sample ($L_{bol} \sim 1.9 \times 10^{41}\, \rm{erg\, s^{-1}}$). Although this LLAGN belongs to the type 2 class, the nucleus is not obscured, as suggested by X-ray ($N_H \sim 10^{21}\, \rm{cm^{-2}}$, \cite{2006A&A...460...45G}) and optical observations ($E(B - V) = 0.08$, \cite{1997ApJS..112..315H}). Fig.~\ref{n4594_sed} shows the high-spatial resolution SED (black dots), together with the measurements based on large apertures (grey spikes). As in the case of NGC~1052, the optical range is dominated by the stellar emission in low-spatial resolution data. At high-spatial resolution, NGC~4594 shows a thermal-like bump with a maximum at $\sim 1\, \rm{\mu m}$. This component presents a soft variability in a timescale of a few years, thus cannot have a stellar origin. On the other hand, this thermal-like bump in the IR can be associated to a truncated accretion disk, expected for LLAGNs \cite{2005Ap&SS.300..177N}, although the non-detection of the Fe\,K line emission in X-rays by \cite{2003ApJ...597..175P} suggests the absence of an accretion disk in this object. Another possible origin for this emission could be associated with synchrotron emission produced in the base of a jet, as in the case of X-Ray binaries in the low-hard state \cite{2005ApJ...635.1203M,2008ApJ...681..905M}.

\section{Average LLAGN at parsec scales}

\begin{figure}
  \centering
  \includegraphics[width = 0.5\textwidth]{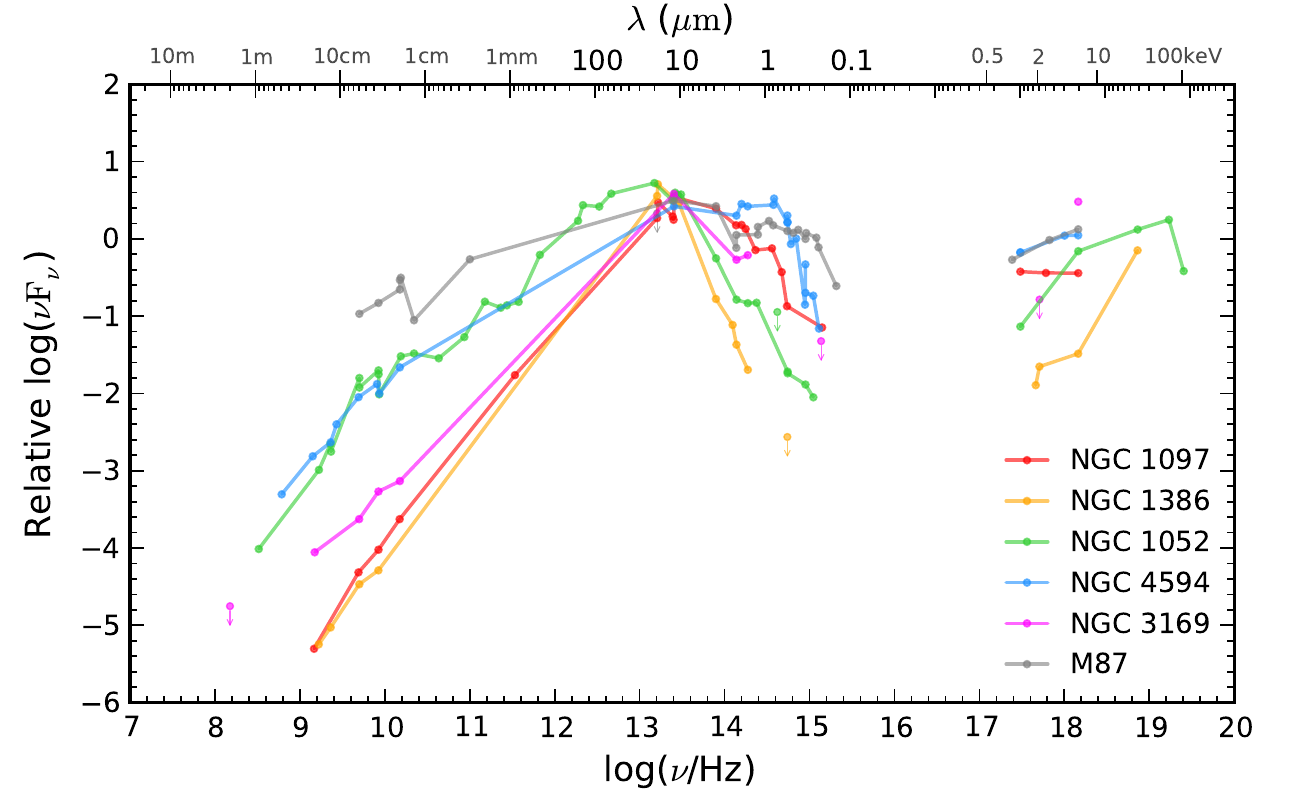}~
  \includegraphics[width = 0.5\textwidth]{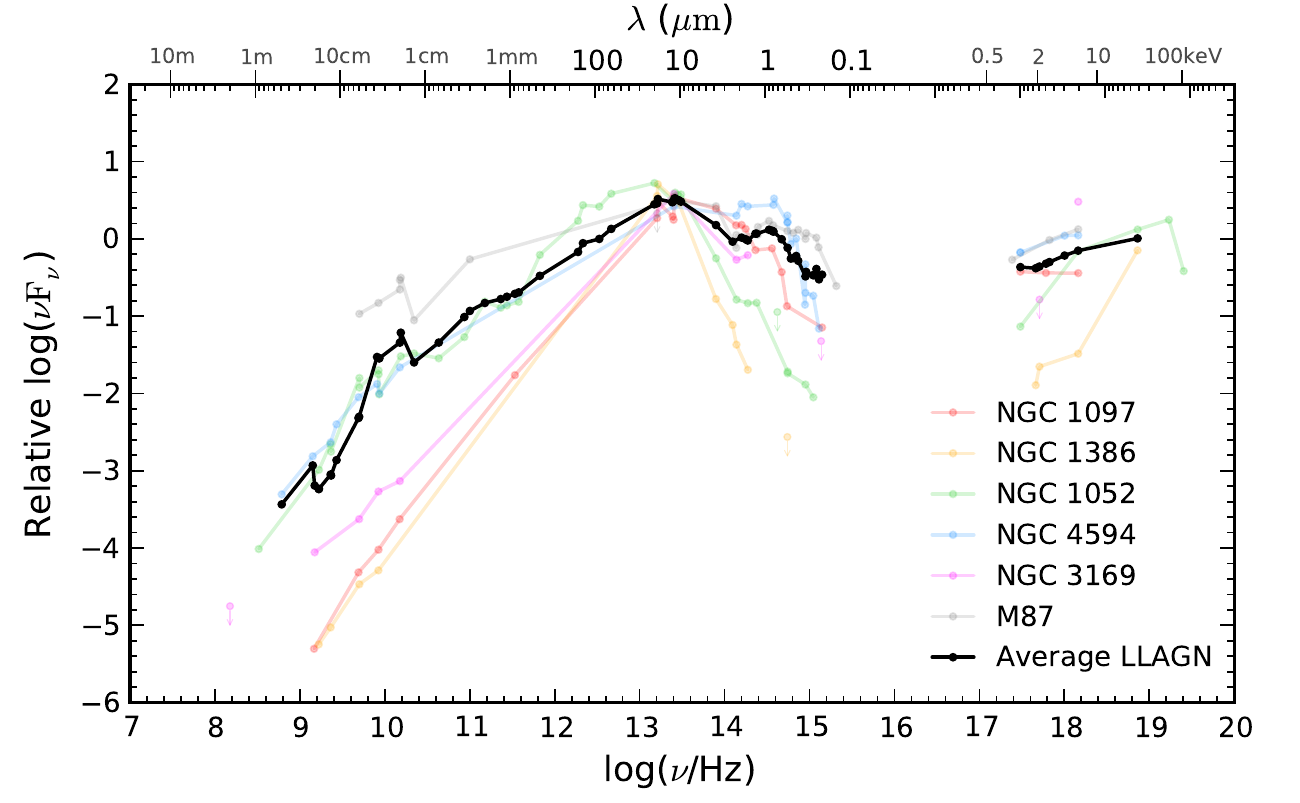}
  \caption{\label{average_sed} \emph{Left panel:} high-spatial resolution SEDs for the six LLAGNs in the sample.\emph{Right panel:} in black, average SED for LLAGNs obtained from the individual SEDs shown in the left panel.}
\end{figure}

Left panel in Fig.~\ref{average_sed} shows the high-spatial resolution SED for all the LLAGNs in the sample. Each SED has been normalized by its mean power ($\overline{\nu F_\nu}$) in order to facilitate their comparison. The dispersion of the SED shapes is large, specially in the mid-IR-to-UV range. However, three main characteristics are common for all the LLAGNs in the sample: \emph{i)} a flatter spectra at radio wavelengths when compared with Seyfert galaxies, associated to the presence of compact cores and jets, \emph{ii)} the bulk of the emission is released in the IR range, and \emph{iii)} the IR luminosity is very similar to the X-ray luminosity. 
The latter suggests that there might be a relation between both ranges. In Seyferts and quasars, the IR emission is produced by dust in the torus, as a consequence of the absorption and re-emission of the optical and X-ray light. In LLAGNs the presence of the torus component is not so clear. Figs~\ref{n1052_sed} and \ref{n4594_sed} show that the IR is not apparently dominated by a thermal component, as in the case of Seyfert nuclei. Notwithstanding, the mechanism that drives the energy output in LLAGNs seems to connect both ranges. 

In order to compare with previous studies, an average LLAGN template was obtained from the individual SEDs. The method described in \cite{2010MNRAS.402..724P} was followed to derive the average LLAGN SED (right panel in Fig.~\ref{average_sed}). The main characteristics mentioned before are reflected in the average template, which shows a flat radio continuum and a maximum in the mid-IR. The diversity of SEDs in the mid-IR-to-UV range translates into a mix, in the average LLAGN, of the main individual features. These are the thermal-like bump around $\sim 1\, \rm{\mu m}$ plus a steep power-law continuum decreasing from the mid-IR to the UV.

\subsection{Comparison with previous studies}
\begin{figure}
  \centering
  \includegraphics[width = 0.5\textwidth]{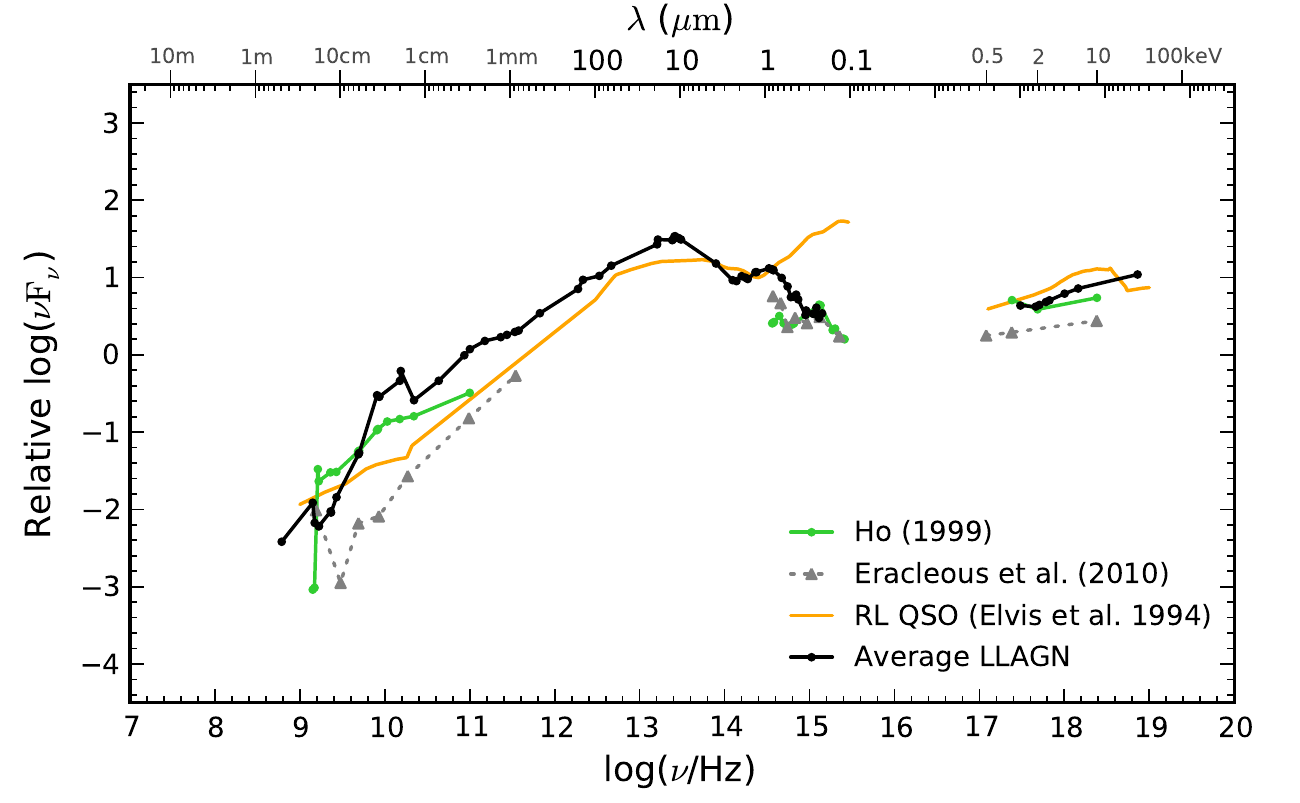}~
  \includegraphics[width = 0.5\textwidth]{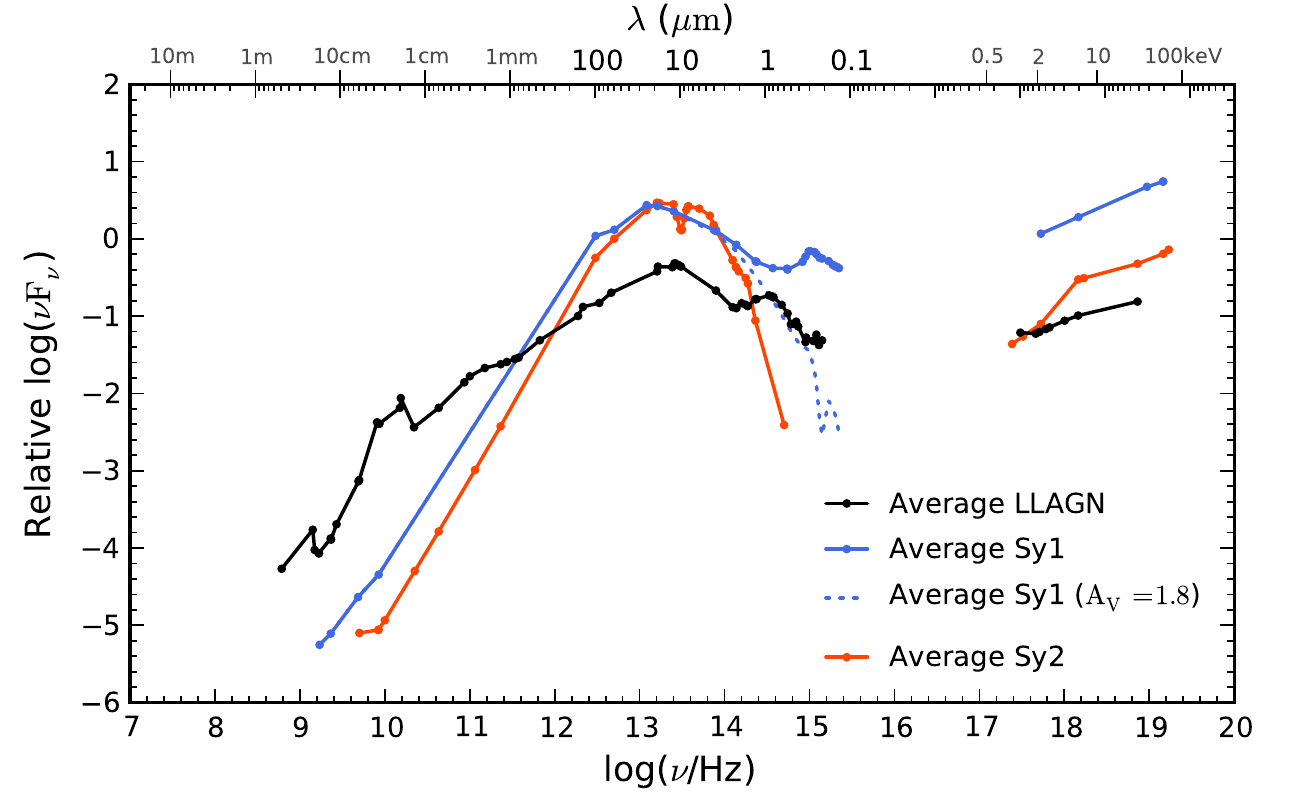}
  \caption{\label{average_Ho_Sy} \emph{Left panel:} comparison of the LLAGN average SED obtained in this study (black dots) with previous studies based on high-spatial resolution data, from \cite{1999ApJ...516..672H} (green dots) and \cite{2010ApJS..187..135E} (grey triangles). In orange, the radio-loud template for QSOs provided by \cite{1994ApJS...95....1E}. \emph{Right panel:} high-spatial resolution average SED for LLAGNs (black dots), Seyfert 1s (blue dots, \cite{2010MNRAS.402..724P}) and Seyfert 2s (red dots, \cite{2010MNRAS.402..724P}).}
\end{figure}

Left panel in Fig.~\ref{average_Ho_Sy} shows the average LLAGN (black dots) together with an average SED distribution obtained from the sample of LINERs in \cite{1999ApJ...516..672H} (green dots), using the same method as in our sample. The average LINER SED given by \cite{2010ApJS..187..135E} is also shown (grey triangles). Overall, our average SED for LLAGNs is consistent with previous works in the common wavelength ranges covered. LLAGNs are bright at radio wavelengths but do not show the big blue bump, footprint of the accretion disk, in the optica/UV range. In contrast, the spectral slope in this part of the spectrum is very steep, decreasing with wavelength. At X-rays, the SED becomes almost flat for all the templates.

However, subarsec resolution data has been included for the first time in the NIR (VLT/NaCo adptive opctics) and mid-IR (VLT/VISIR diffraction limited data) ranges. This permit us to characterize the region of the spectrum in which these faint nuclei radiate most of their luminosity. As mentioned before, the nature of the emission in this range seems to be associated with non-thermal processes. For comparison, the average SED for ``radio-loud'' quasars given by \cite{1994ApJS...95....1E} is also shown in the left panel of Fig.~\ref{average_Ho_Sy} (orange line). As pointed out by previous authors, the ``radio-loud'' quasar template presents also a similar shape when compared with LLAGNs in the radio and X-ray ranges.

\subsection{Low and high-luminosity AGNs}
The average SED for Seyfert 1 (in blue) and Seyfert 2 nuclei (in red) derived from high-spatial resolution data by \cite{2010MNRAS.402..724P} are shown in the right panel of Fig.~\ref{average_Ho_Sy}. Compared with the LLAGN template, Seyferts nuclei exhibit a strong and thermal mid-IR component, associated with the emission from the dust in the torus. At radio wavelengths, both Seyfert SEDs show a steeper decrease, more similar to ``radio-quiet'' quasars, in contrast with LLAGNs which are ``radio loud''. In the optical/UV range, the differences between both classes of AGN are larger. Seyfert 1s present a big blue bump, i.e. the emission from the accretion disk, although this is softer when compared with the one in quasars \cite{2010MNRAS.402..724P}. In contrast, the shape of the Seyfert 2 template is determined by the obscuration in these nuclei. This is in line with the Unified Model, in which the torus hides the central engine for type 2 objects. However, LLAGNs present a significantly different shape when compared with Seyferts. Furthermore, its SED cannot be reproduced by applying a foreground extinction to the Seyfert 1 template (dashed line in left panel of Fig.~\ref{average_Ho_Sy}).

\section{Summary}

The importance of LLAGNs lies in the fact that they permit to explore the limits of the classical picture for active nuclei. At very low luminosities ($L_{bol} \lesssim 10^{42}\, \rm{erg\, s^{-1}}$) the broad-line region and the torus are expected to vanish, giving way to a radiatively inefficient structure. As a consequence of the changes in the inner structure, the SED of LLAGNs is expected to present differences with regard to the bright class.

A high-spatial resolution study has been performed for a sample of 6 nearby LLAGNs. This includes sub-arcsec resolution data in the radio, IR, optical, UV and X-ray ranges and allows us to disentangle the nuclear light from that of the host galaxy. For the first time, the mid-IR to NIR range is sampled at high-spatial resolution, a range in which LLAGNs radiate most of their luminosity. The shape of the SED in LLAGNs suggests that non-thermal processes can dominate the emission from radio to UV wavelengths. Self-absorbed synchrotron emission in a jet is a possible mechanism to explain the radio-to-UV continuum \cite{2008ApJ...681..905M}. Still, in the case of the Sombrero galaxy this synchrotron continuum is not present in the IR-to-UV range, and the thermal-like emission could be associated to a truncated accretion disk.

Overall, the high-spatial resolution SEDs of the LLAGNs in the sample are intrinsically different when compared with bright AGNs. The big blue bump is absent, even for unobscured nuclei. The strong thermal IR component associated with the torus in Seyfert nuclei is not detected for LLAGNs, in line with numerical simulations that predicts this structure to receed at low-luminosities. If it is still present, the torus does not seem to dominate the emission in faint nuclei. Nevertheless, there are still some similarities with the brighter counterparts. At radio wavelengths, LLAGNs show a similar spectral index as ``radio-loud'' quasars. Moreover, the behaviour of faint Seyferts (e.g. NGC~1386) is more similar to that of LLAGNs rather than the case of bright Seyferts, suggesting a smooth transition between both classes of AGNs.


\section*{References}
\bibliographystyle{iopart-num.bst}
\bibliography{agn}

\end{document}